\documentclass[%
 reprint,
superscriptaddress,
 amsmath,amssymb,
 aps,
pra,
]{revtex4-2}
\usepackage{multirow}
\usepackage{graphicx}
\usepackage{babel}
\usepackage{color}
\usepackage{dcolumn}
\usepackage{bm}
\usepackage{comment}

\usepackage[normalem]{ulem}
\usepackage{url}
\usepackage[allcolors=black,colorlinks,breaklinks]{hyperref}

\def\R{\mathbb{R}}

\def\e{\text{e}}

\newcommand{\rbf}{\mathbf{r}}
\newcommand{\xbf}{\rbf}
\newcommand{\pos}[1][]{{\mathbf{q}_{#1}}}
\newcommand{\mom}[1][]{{\mathbf{p}_{#1}}}
\newcommand{\wm}[1][]{{\mathcal{C}_{#1}}}

\newcommand{\zbf}[1][]{{\mathbf{z}_{#1}}}
\newcommand{\nf}[1][]{{\gamma_{#1}}}
\newcommand{\gwp}[1][]{{g_{#1}}}
\newcommand{\sgwp}[1][]{{u_{\text{mfr}}}}

\usepackage[dvipsnames]{xcolor}

\begin{document}

\newcommand{\eqstar}{\textsuperscript{*}}

\preprint{APS/123-QED}

\title{Meshfree \textit{versus} grid-based Schrödinger solvers for modeling the interactions between free-electron wave packets and light}

\author{Mitja Funk }

    \email{funk@physik.uni-kiel.de}
    \affiliation{%
Institute for Experimental and Applied Physics, Kiel University, Leibnizstraße 19, 24118 Kiel, Germany
}%

\author{Sebastian Merk}

    \email{s.merk@tum.de}

    \affiliation{Department of Mathematics, Technical University of Munich,  Boltzmannstraße 3, 85748 Garching bei München, Germany}

\author{Caroline Lasser}
    \email{classer@tum.de}
    \affiliation{Department of Mathematics, Technical University of Munich,  Boltzmannstraße 3, 85748 Garching bei München, Germany}

\author{Marlis Hochbruck}
    \email{marlis.hochbruck@kit.edu}
    \affiliation{Institute for Applied and Numerical Mathematics, Karlsruhe Institute of Technology, Englerstraße 2, 76131 Karlsruhe,
Germany}

\author{Nahid Talebi}%
 \email{talebi@physik.uni-kiel.de}
\affiliation{%
Institute for Experimental and Applied Physics, Kiel University, Leibnizstraße 19, 24118 Kiel, Germany
}%

\date{\today}

\begin{abstract}


The interaction of free-electron wave packets with electromagnetic fields provides a powerful route toward coherent electron control, enabling the generation of energy combs, momentum-state superpositions, and aberration-engineered electron beams. Existing theoretical descriptions, however, often rely on eikonal or no-recoil approximations. Here, we present a mesh-free numerical framework that directly solves the time-dependent single-particle Schrödinger equation for arbitrary electromagnetic potentials. Comparison with a benchmark mesh-based Schrödinger solver reveals excellent quantitative agreement. By eliminating the need for spatial meshing, our method offers an efficient and scalable route for simulating electron wave packet dynamics in complex time-dependent and static electromagnetic environments, while the simulation time is significantly improved by up to 800 times faster. These capabilities establish a versatile computational tool for quantum electron optics and free-electron-light interactions beyond eikonal approximations.  

\end{abstract}

\maketitle


\section{Introduction}\label{sec:intro} 

Propagating electron pulses in ultrafast electron microscopes constitute highly controllable quantum wave packets, whose temporal duration, spatial profile, and interactions with matter and electromagnetic fields can nowadays be engineered with remarkable precision within the framework of ultrafast electron microscopy \cite{Zewail2010,Barwick2009-xk,Feist2015-xs,Tsesses2023-hd,Morimoto2018,Vanacore2018,Kfir2020,Hassan2017}. In parallel, the emerging field of quantum nanophotonics with free electrons \cite{Garcia_de_Abajo2025-lh,Ruimy2025-vk,Talebi2026, Taleb2023, Taleb2025} aims to tailor the spatio-temporal structure of free-electron wave packets and their interactions with light and matter at the quantum level, enabling the investigation of quasi-particle dynamics and material excitations with nanometer spatial and femtosecond temporal resolutions. These developments require theoretical and computational frameworks capable of accurately describing the dynamics of electron–light interactions to precisely model the change in the shape of the electron wave packets while interacting with complex photonic environment.\\

Interactions between free electrons and electromagnetic fields are often described semi-analytically using scattering theory within the no-recoil approximation \cite{Park_2010,Abajo2010,DiGiulio2019}. While highly successful in many regimes, such approaches cannot fully capture the dynamical evolution of the electron wave packet during the interaction process. In particular, they become insufficient in regimes where electrons recoil, strong-field effects, or dynamics beyond the rotating-wave approximation lead to phenomena such as quantum-path interference between single and two-photon interaction mechanisms \cite{Talebi_2019} and near-field-mediated Kapitza–Dirac effects \cite{Talebi2020,Chahshouri31122026}. These limitations have motivated the development of numerical approaches based on the minimal-coupling Hamiltonian, enabling a more complete treatment of coupled electron–field dynamics \cite{Talebi2016, Talebi2017}.\\

An approach to numerically simulate the interactions within the minimal-coupling Hamiltonian is to combine a time-dependent grid-based Schrödinger solver with a time-dependent Maxwell solver. Such approaches generally require multiscale treatments due to the substantial mismatch between the spatial and temporal discretizations scales associated with the Schrödinger and Maxwell equations. In this context, hybrid frameworks based on pseudospectral methods for the Schrödinger equation together with finite-difference time-domain (FDTD) solvers for Maxwell’s equations have been successfully applied to various near-field \cite{Chahshouri2023,Chahshouri2025} and free-space interaction geometries \cite{Ebel2025,Ebel2023}.  

The time and spatial scales clearly show that the Schrödinger solver is the bottleneck within the coupling with the Maxwell solver. Hence, here, we demonstrate a mesh-free method for the Schrödinger equation with the minimal-coupling Hamiltonian based on thawed Gaussian wave packet approximation (TGWP), and benchmark it with the grid-based scheme stated above. We apply both methods to two cases where the electromagnetic field is prescribed analytically, and closely compare the results obtained using the mesh-free method versus the grid-based Schrödinger solver.


\begin{figure*}
\includegraphics[width=1\linewidth]{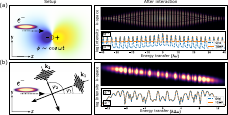}
\caption{\label{fig:setup} We consider electron-light interactions in terms of dipole coupling (a) and stimulated Compton scattering (b). The central quantity of comparison is the momentum space distribution of the electron after interaction and the resulting longitudinal intensity (energy transfer). The grid-based reference solution is in blue, the mesh-free approximation, the thawed Gaussian wave packet approximation (TGWP) is in orange. For Compton scattering (b), the agreement is excellent, for dipole coupling~(a) the numerical noise is of the order of experimental resolution.} 
\end{figure*}
\section{Model Systems}\label{sec:models} 

We model two complementary examples of electron-light interactions. The quasi-static near-field describes localized nanostructure-mediated coupling with highly inhomogeneous electromagnetic fields, whereas stimulated Compton scattering represents a free-space interaction in which coherent absorption and stimulated emission from two laser fields produce a well-defined momentum transfer governed by the optical wave-vector difference.

We consider a time-dependent Schrödinger equation 
\begin{subequations} \label{eq:schroedi}
    \begin{equation} \label{eq:schroedi-a}
        i\hbar\partial_t\psi(t,\xbf) = \widehat{H}\psi(t,\xbf), \quad \psi(0,\xbf) = \psi_0(\xbf)
    \end{equation}
    for a two-dimensional configuration $\xbf= x \mathbf{e}_x + y \mathbf{e}_y$, and with the Hamiltonian specified as
    \begin{equation}   \label{eq:schroedi-b}
    \widehat{H}= -\frac{\hbar^2}{2m}\nabla^2+ V(t,\xbf).
    \end{equation}
The initial electron wave packet is considered as $\psi_0(\xbf)=\gwp(\xbf)$, that is a 
Gaussian wave packet 
\begin{equation*}
    \gwp(\xbf) = \nf_0 \exp\!\left(\frac{i}{\hbar}\bigl(\frac{1}{2}(\xbf-\pos)^\top \wm(\xbf-\pos) + \mom^\top (\xbf-\pos)\bigr)\right).
\end{equation*}
with parameters $(\pos_0,\mom_0,\wm_0,\nf_0)$.
The position and the momentum center are denoted by $\pos$ and $\mom$, respectively. The width matrix $\wm$ is a complex symmetric matrix with a positive definite imaginary part, so that the wave packet is square integrable. For our initial data, the width matrix is of the form $\wm_0 = i\Gamma_0$ with a real symmetric positive definite matrix $\Gamma_0$. 
The complex number $\nf_0$ accounts for normalization and phase factors. For our experiments, the initially Gaussian-shaped wave packet has spatial widths (FWHM) $\sigma_x = 400\,\mathrm{nm}$ and $\sigma_y = 80\,\mathrm{nm}$, corresponding to $\Gamma_0 = \hbar/2\,\mathrm{diag}(1/\sigma_x^2,\, 1/\sigma_y^2)$. Before interaction, the initial velocity points solely in the
$x$-direction and is given by the relativistic dispersion relation for the kinetic energy $U_e$ (cf. Tab.~\ref{tab:directcomparison_dip} and Tab.~\ref{tab:directcomparison_scs}).

\end{subequations}

\subsection{Quasi-static near-field}

The first model case that we consider is the interaction of a free-electron wave packet with the optical near-field generated by a gold nanorod, representing a prototypical configuration for photon-induced near-field electron microscopy (PINEM) \cite{Barwick2009-xk,Feist2015-xs}. Generally, for simulating the response of such a nanorod in interaction with an incident pulsed laser excitation, a Maxwell solver has to be employed. Here, however, we employ an analytical quasi-static dipole model \cite{Hergert_2021} in order to isolate and benchmark the performance of both Schrödinger propagation schemes independently of the accuracy of the Maxwell's solver.

The interaction is described by the time-dependent potential
\begin{equation}
    V(t,\mathbf{r}) = -e\phi_0(\mathbf{r})\cos(\omega t ),
\end{equation}
with
\begin{align}
\phi_0(\mathbf{r})=
\begin{cases}
E_0 x \left|\dfrac{\varepsilon_G - 1}{\varepsilon_G + 1}\right|, & r < R, \\
E_0 x \dfrac{R^2}{r^2}\left|\dfrac{\varepsilon_G - 1}{\varepsilon_G + 1}\right|, & r \ge R.
\end{cases}
\end{align}
where $R$ denotes the nanorod radius, $E_0$ the incident electric field amplitude, and $\varepsilon_G$ the dielectric function of gold. The rod is oriented out of plane with its center displaced by $\sigma_y/2$ perpendicular to the electron trajectory, such that the electron interacts with the structure in an aloof configuration (cf. Fig.~\ref{fig:setup}). The initial electron position is chosen sufficiently far from the interaction region to ensure negligible coupling at $t=0$. Specifically, the wave packet is initialized $1.5\,\mathrm{\mu m}$ from the center of the nanorod along the $x$-axis.

\subsection{Stimulated Compton scattering}

While PINEM relies on nanostructures to mediate the interaction between free electrons and optical fields, electron-light interactions can also occur directly in free space. The near-field-mediated energy transfer between the light field and electron wave packets is dominated by the direct single-photon absorption and emission process, whereas the free-space interaction includes two-photon processes. One well-known example is the Kapitza-Dirac effect, in which an electron wave packet is diffracted by the standing-wave pattern formed by two counter-propagating laser beams of equal wave length \cite{Kapitza_Dirac_1933,Freimund2001,Lin2024,Moriova2025}. A generalization of this process is the stimulated Compton scattering, where the wave packet interacts with two superposed inclined laser pulses of differing wave lengths \cite{, Kozak2018, Chahshouri31122026, Ebel2025} (cf. Fig.~\ref{fig:setup} (b)). 

During this interaction process, energy and momentum are exchanged through absorption and stimulated emission processes from both laser pulses, resulting in a coherent momentum transfer determined by the difference of the laser wave vectors. In contrast to the localized near-field interaction discussed above, this mechanism predominantly induces a well-defined and unique momentum transfer and gives rise to highly structured interference patterns in the momentum space.

The vector potential describing this interaction is the superposition of two pulsed Gaussian beams 
\begin{equation}
    \mathbf{A}(t,\mathbf r)=\mathbf{A}^{(1)}(t,\mathbf r)+\mathbf{A}^{(2)}(t,\mathbf r).
\end{equation}
Each contribution $\mathbf{A}^{(i)}$ is modeled as a spatially-structured paraxial beam, whereas temporally the pulse is following the formulation of Porras \cite{PhysRevE.58.1086}, with pulse duration $\delta t = 20\,\mathrm{fs}$ and waist width $a_0 = 1400\,\mathrm{nm}$ (see Supplemental Material). The polarization of the vector potential is chosen along the out-of-plane direction, such that the effective interaction is governed by the ponderomotive potential
\begin{equation}
V(t,\mathbf{r}) = \frac{e^2}{2m}|A_z(t,\mathbf{r})|^2,
\label{v_pond}
\end{equation}
where $A_z(\mathbf{r},t)$ denotes the out-of-plane component of the vector potential.

Efficient momentum transfer requires the fulfillment of the phase-matching condition and therefore ensuring simultaneous conservation of energy and momentum during the interaction \cite{Hall1963,PhysRevLett.14.851,Kozak2018,Chahshouri31122026}. For the two-dimensional geometry considered here, the phase-matching condition is given by
\begin{equation}
\Delta\omega = \omega_1 - \omega_2 = \frac{v_0}{c}(\omega_1\cos\varphi_1-\omega_2\cos\varphi_2),
\label{eq_phase_matching}
\end{equation}
with $\omega_i$ denoting the laser frequencies, $\varphi_i$ the inclination angles of the beams, and $v_0$ the electron velocity. We use a class of parameters for the electron wave packet and light pulses that satisfy this condition (cf. Tab.~\ref{tab:directcomparison_scs}) and also investigate the resulting interaction behavior when the parameters deviate from the idealized case, using both the mesh-free and the grid-based simulation schemes. The electron is initialized $1\,\mathrm{\mu m}$ from the laser-pulse crossing point.


\section{Computational Methods} \label{sec:methods}

We compare two different numerical methods for solving the time-dependent Schrödinger equation \eqref{eq:schroedi}.
As the grid-based reference, we apply a split-step Fourier method using Strang splitting \cite{Lub08,strang_split_step}. This widely established Fourier-based method has been effectively used in the past to explore the dynamics of electron wave packets interacting with complex time-dependent electromagnetic fields \cite{Talebi2016,Talebi2020}.

The mesh-free approximation 
relies on a sum of Gaussian wave packets
\begin{equation}\label{eq:ansatz}
\psi(t,\xbf) \approx \sum_{j=1}^N c_j\, \e^{i S_j(t)/\hbar} \,\gwp[j](t,\xbf)
\end{equation}
with time-independent complex coefficients $c_j$ and time-dependent Gaussian wave packets 
$\gwp[j](t,\xbf)$. Each wave packet is determined by its parameters, the  
center in position $\pos[j](t)$ and momentum $\mom[j](t)$, the 
width matrix $\wm[j](t)$, and the normalization and phase factor $\nf_j(t)$. The Fourier transform of a complex Gaussian wave packet has a closed form \cite[Appendix A]{Fol89}, enabling mesh-free computation of the momentum distribution.

\subsection{Initial mesh-free representation}
For an arbitrary but fixed width matrix $\wm$, a family of Gaussian functions forms an over-complete set, providing 
    \begin{equation}\label{eq:int}
        \psi_0 = (2\pi\hbar)^{-d} \int_{\R^{2d}} \langle \gwp[\zbf]\mid \psi_0\rangle \,\gwp[\zbf]\, d\zbf,
    \end{equation}
where the integration is with respect to $\zbf = (\pos, \mom)$. As part of the computational method, one chooses 
the width matrix $\wm = i\Gamma$ of the the representing Gaussian functions $\gwp[\zbf]$, where $\Gamma$ is a suitable real symmetric, positive definite matrix. For the initial Gaussian $\psi_0$, there is an explicit formula for the overlap $\langle \gwp[\zbf]\mid \psi_0\rangle$, see, e.g., \cite[Lemma~1]{BerL22}, which motivates to introduce the Gaussian weighting function 
\[
    W(\zbf) = \exp\left(-\frac{1}{2\hbar}(\zbf-\zbf[0])^\top\Sigma_0^{-1}(\zbf-\zbf[0])\right),
\]
that is centered in $\zbf_0 = (\pos_0,\mom_0)$ and has inverse width matrix $\Sigma_0$ with
\[
\Sigma_0 = \begin{pmatrix}\Gamma_0^{-1} + \Gamma^{-1} & 0\\ 0 & \Gamma_0 + \Gamma\end{pmatrix}.
\]
For quasi-Monte Carlo quadrature with weight $W(\zbf)$, one constructs normally distributed Sobol' points 
$\zbf_1,\ldots,\zbf_N$ according to $W(\zbf)$ and approximates the integral \eqref{eq:int} as
    \[
    \psi_0 \approx  \frac{1}{N (2\pi\hbar)^{d}} \sum_{j=1}^N  \frac{\langle \gwp[\zbf_j]\mid \psi_0\rangle}{W(\zbf_j)} \,\gwp[\zbf_j]
    \]
This determines the linear expansion coefficients of the wave packet ansatz \eqref{eq:ansatz} as
\[
c_j = \frac{\langle \gwp[\zbf_j]\mid \psi_0\rangle}{N (2\pi\hbar)^{d} W(\zbf_j)}.
\]
\subsection{Approximate Evolution}

\begin{figure*}
\includegraphics[width=1\linewidth]{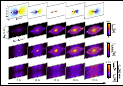}
\caption{\label{fig:time_ev_dip} Snapshots of the propagating wave packet interacting with the quasi-static dipole field for different selected times  for the parameter set $\{E_0 = 0.5\,\mathrm{V/nm}; U_e=20\,\mathrm{keV} \}$ . The top row shows the position in real space, the middle rows show the absolute wave function during interaction for the two methods. The bottom row shows the regularized error. The mesh free approximation resolves the dynamics in regions of high intensity with rich dynamics in both transverse and longitudinal direction.  In the regions of negligible intensity, the error is dominated by numerical noise and relatively high. The max value for the top two colorbars is set to $\max\big(|\psi_k^\mathrm{grid}|\big)/150 \approx 8.75\times10^{-10} \,\mathrm{m}$.}
\end{figure*}

By linearity of the Schrödinger equation \eqref{eq:schroedi}, we evolve the Gaussians separately and semi-classically. 
We use the Hagedorn parametrization of the width matrices $\wm_j(t) = P_j(t)Q_j(t)^{-1}$, see \cite{LasL20}, and solve the classical equations of motion
 \begin{align*}
        &\dot{\pos}_j(t) = \frac1m \mom_j(t), \quad\dot{\mom}_j(t) = -\nabla V(t,\pos_j(t)),\\
        &\dot{Q}_j(t) = \frac1m P_j(t),\quad \dot{P}_j(t) = -\nabla^2 V(t,\pos_j(t)) Q_j(t),\\
        &\dot{S}_j(t) = \tfrac{1}{2m }\mom_j(t)^\top \mom_j(t) - V(t,\pos_j(t)),
    \end{align*}
subject to the initial conditions 
    \begin{align*}
        &(\pos_j(0), \mom_j(0)) = \zbf_j,\\
        &Q_j(0) = \sqrt\hbar\Gamma^{-1/2},\ P_j(0) = i\sqrt\hbar\Gamma^{1/2},\ S_j(0) = 0.
    \end{align*}
Note that the normalization factor $\gamma_j(t)$ is complex and depends on the determinant of $Q_j(t)$, see \cite{LasL20}.     
This way, each Gaussian is moving in its own effective potential, that is the local harmonic approximation of $V(t,\xbf)$ around its position center,    
\begin{align*}
        V_{\mathrm{eff},j}(t,\xbf) &= V(t,\pos_j(t)) + (x-\pos_j(t))^\top \nabla V(t,\pos_j(t)) \\
        &+ \frac{1}{2}(\xbf-\pos_j(t))^\top \nabla^2 V(t,\pos_j(t)) (\xbf-\pos_j(t)).
    \end{align*}    
The variational equations of motion provide a more accurate approximation \cite{BurDHL24} than the purely classical one employed here, but are more costly due to the variational averaging of the potentials. Here, we have solved  
    the parameter equations with a Störmer-Verlet integrator for order 2.

\section{\label{sec3} Results \& Discussion}
We treat the grid-based solution as the reference against which the TGWP approximation is evaluated. Qualitatively, we compare the modulus of the wave function and the longitudinal energy transfer during and after the interaction. Quantitatively, we assess the  agreement using $L^2$ errors and error measures specifically designed to characterize the sideband structure of the longitudinal marginal of the momentum distribution.

For the grid-based reference, the computational box has dimensions $L_x \times L_y = 8\,\mu\mathrm{m} \times 5\,\mu\mathrm{m}$, and the numerical parameters are chosen such that the solution is sufficiently self-converged while keeping the computation time within reasonable bounds (cf. Tab.~\ref{tab:num_params}).

To enable a fair comparison of computational performance, the time step for the TGWP approximation is chosen such that its self-convergence with respect to the time step is comparable to that of the reference solution, and the number of Gaussians is chosen such that the $L^2$ error between the two solutions has saturated.

Tab.~\ref{tab:num_params} summarizes the numerical parameters used for both the grid-based reference and the TGWP approximation.

\begin{table}[htbp]
\centering
\renewcommand{\arraystretch}{1.3}
\begin{tabular}{|l|cc|ccc|}
\hline
 & \multicolumn{2}{c|}{Grid-solver} & \multicolumn{3}{c|}{TGWP} \\
\cline{2-3} \cline{4-6}
 & $\Delta x, \Delta y$ & $\Delta t$ & Gaussians & $\Delta t$ & $\Gamma$ \\
\hline
II.A & $2.2\,\mathrm{nm}$ & $0.01\,\mathrm{fs}$ & $2^{18}$ & $0.01\,\mathrm{fs}$ & $100^2 \Gamma_0$ \\
II.B & $2.2\,\mathrm{nm}$ & $0.01\,\mathrm{fs}$ & $2^{15}$ & $0.3\,\mathrm{fs}$ & $50\, \mathrm{diag}(\hbar/\sigma_y^2,\, \hbar/\sigma_y^2)$ \\
\hline
\end{tabular}
\caption{Numerical parameters for the grid-based reference and the TGWP approximation. For the split-step, the grid spacing and time step are given. For the TGWP approximation, the number of Gaussians, the time step, and the choice of initial width matrix $\Gamma$ are listed.}
\label{tab:num_params}
\end{table}

A detailed assessment of the convergence behavior is provided in the Supplement.

\begin{figure*}
\includegraphics[width=1\linewidth]{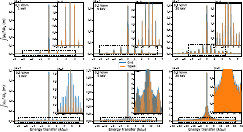}
\caption{\label{fig:marg_after_int_dip} Longitudinal energy transfer spectra for all considered combinations of parameters for the quasi-static near-field. The agreement of the TGWP with the grid-based method decreases with decreasing electron kinetic energy and with increasing electric field strength. } 
\end{figure*}

\subsection{Quasi-static near-field}

\begin{figure*}
\includegraphics[width=1\linewidth]{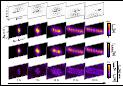}
\caption{\label{fig:time_ev_compton} Time evolution of the electron wave packet interacting with two laser pulses via stimulated Compton scattering shown for the parameter set \{ $\lambda_1 = 474.31\,\mathrm{nm};\,\lambda_2 = 400\,\mathrm{nm};\, E_0=20\,\mathrm{V/nm} $ \}. The top row shows the electron wave packet and the $z$-component of the vector potential in real space at depicted propagation times. Max wave function error is 0.988\%. The max value for the top two colorbars is set to $\max\big(|\psi_k^\mathrm{grid}|\big)/5 \approx 1.4\times10^{-8} \,\mathrm{m}$.} 
\end{figure*}

\begin{figure*}
\includegraphics[width=1\linewidth]{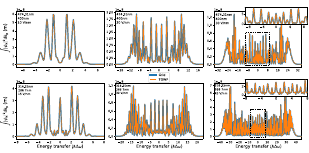}
\caption{\label{fig:comparison_compt} Longitudinal energy transfer spectra for all considered combinations of parameters for stimulated Compton scattering. The overall very good agreement decreases with increasing electric fields strength. } 
\end{figure*}

We first investigate the interaction of a free-electron wave packet with the localized near-field generated by an oscillating dipole, as illustrated in Fig.~\ref{fig:setup} (a). Explicit form of the potential is provided in section II.A. We deploy different parameter combinations of electric field amplitude $E_0$ and electron kinetic energy $U_e$ listed in Tab.~\ref{tab:directcomparison_dip}. 

The strong interaction between the electron wave packet and the optical near-field leads to a coherent energy transfer from the optical near-field excitation to the electron wave packet. This effect leads to an appearance of an energy comb with peaks forming at exactly $\hbar k_e\pm n\hbar k_{ph}$ along the longitudinal direction, where $k_{ph}$ is the wave number of the photon.
This effect has been demonstrated for various configurations of near-field excitation, regardless of the shape of the nanostructure used or the laser excitation scheme \cite{Dahan2020,Wang2020,Kfir2020}. Hence, the near-field configuration only mediates the interaction between light and electron beams. However, the strength of the interaction strongly depends on different parameters including the excitation wavelength, the electron velocity, the impact parameter, and the spatial configuration of the near-field, since the latter determines the momentum distribution of the near-field excitation in the reciprocal space.\\
In addition to the longitudinal momentum transfer, simulations have already predicted a transverse momentum recoil and energy-dependent diffraction effect \cite{Talebi2020}. This leads to a transverse modulation of the electron wave packet akin to the Kapitza-Dirac effect, however at larger momentum orders, due to the large momentum component of the evanescent near-field distribution.

\begin{table*}
\centering
\small
\renewcommand{\arraystretch}{1.15}

\begin{ruledtabular}
\begin{tabular}{ccccccccc}
$\lambda$ (nm) &
$E_0$ (V/nm) &
$U_e$ (keV) &
$T_\mathrm{prop}$ (fs) &
$\epsilon_{L^2}$ &
$\epsilon^\mathrm{abs}_{L^2}$ &
missed peaks (\%) &
height err. (\%) &
width err. (\%) \\
\hline
\multirow{6}{*}{800}
& \multirow{3}{*}{0.1}
& 1  & 160 & 0.195 & 0.155 & --- & 0.754 & 0.00 \\
& & 5  & 100 & 0.174 & 0.111 & --- & 0.508 & 4.444 \\
& & 20 & 50  & 0.19 & 0.0872 & --- & 0.174 & 1.176 \\
\cline{2-9}
& \multirow{3}{*}{0.5}
& 1  & 160 & 0.5 & 0.289 & --- & 0.982 & 1.18 \\
& & 5  & 100 & 0.625 & 0.186 & --- & 0.236 & 1.29 \\
& & 20 & 50  & 0.759 & 0.170 & --- & 0.118 & 0.274 \\
\end{tabular}
\end{ruledtabular}

\caption{
Physical parameter sets used for the quasi-static near-field and the related errors.
$\lambda$ denotes the center wavelength of the near-field plasmonic oscillations induced in the nanorod,
$E_0$ the electric-field strength,
$U_e$ the acceleration voltage
and $T_\mathrm{prop}$ the propagation time, chosen sufficiently large such that the electron no longer interacts with the optical field at $t=T_\mathrm{prop}$. 
The dielectric function of gold at $\lambda={800}$ nm is set to 
$\varepsilon_G=-24.061+1.5068i$, the nanorod radius is fixed at 50 nm. For missed peaks --- indicates that no peaks were missed. Note that the $L^2$ errors contain the self-convergence error of the reference solution (cf. Supplemental Material), which grows monotonously with field strength and electron velocity since the convergence parameters are fixed throughout all physical test parameters.}
\label{tab:directcomparison_dip}
\end{table*}

Fig.~\ref{fig:time_ev_dip} shows the time evolution of the electron wave packet during interaction with the dipolar resonance. The split-step solution and the TGWP approximation exhibit close agreement throughout the propagation. In particular, the TGWP method reproduces the transient deformation of the wave packet in the vicinity of the nanostructure as well as the final momentum space distribution after interaction.

While the peaks are reproduced very accurately (cf. Fig.~\ref{fig:time_ev_dip}), the mesh-free method produces artificial oscillations in regions where the wave function almost vanishes. This noise is the main contribution to the error.

The resulting longitudinal energy transfer distributions after the interaction are displayed in Fig.~\ref{fig:marg_after_int_dip} for different combinations of field strength and electron kinetic energy, calculated by integrating the probability distribution in the momentum representation along the transverse axis. Both numerical approaches reproduce the characteristic PINEM sideband structure associated with coherent photon exchange with the localized optical near-field. The agreement regarding spectral envelope, sideband positions, and the overall redistribution of intensity remains very close as the interaction strength increases and the spectrum broadens over a larger energy range.

Table~\ref{tab:directcomparison_dip} reports the error of the full wave function $\epsilon_{L^2}$ as well as of its absolute values $\epsilon_{L^2}^\mathrm{abs}$, both with respect to the $L^2$ norm, which isolates amplitude errors from phase errors. To quantify the sideband structure independently of the background noise, we also show error measures that compare the local maxima of the longitudinal momentum marginal against the reference. For each reference peak larger than $0.1\%$ of the dominant peak, we record whether the TGWP approximation exhibits a corresponding local maximum, and for matched peaks report the relative error in height and full-width-at-half-maximum, normalized to the dominant peak. Across all parameter sets, height errors remain below $1.2\%$ and width errors below $2.3\%$, with at most $1.5\%$ of peaks missed at the strongest interaction.

On the same hardware using 48 CPU threads with largely comparably converged parameters, the computational speedup of the TGWP implementation in \textsc{Julia} compared the split-step implementation in \textsc{MATLAB} is $\approx 100$ across the parameter sets.

\subsection{Stimulated Compton scattering}

\begin{table}
\centering
\small
\renewcommand{\arraystretch}{1.15}

\begin{ruledtabular}
\begin{tabular}{ccccc}
$\lambda_1$ (nm) &
$\lambda_2$ (nm) &
$E_0$ (V/nm) &
$\epsilon_{L^2}$  & $\epsilon^\mathrm{abs}_{L^2}$\\
\hline

 \multirow{3}{*}{474.31}
&   & 10 & 0.045 & 0.018\\
& 400 & 20  & 0.158 & 0.052 \\
& & 30 & 0.342 & 0.09   \\
\cline{1-5}

 \multirow{3}{*}{316.25} 
&   & 15 & 0.087 & 0.038 \\
& 266.7& 30  & 0.314& 0.083 \\
& & 45 & 0.675 & 0.129 \\
\end{tabular}
\end{ruledtabular}
\caption{ Parameter sets and $L^2$ errors for stimulated Compton scattering. To satisfy the phase-matching condition \eqref{eq_phase_matching}, the inclination angles are fixed at $\varphi_1=75^\circ$ and $\varphi_2=105^\circ$ and the electron kinetic energy is set to $30\,\mathrm{keV}$. Consequently, the propagation is also fixed and set to $35\,\mathrm{fs}$. Note that the $L^2$ errors contain the self-convergence error of the reference solution (cf. Supplemental Material), which grows monotonously with field strength and electron velocity since the convergence parameters are fixed throughout all physical test parameters. The computation time amounts to $\approx 1$s for the TGWP method and $\approx 14$min for the split-step method for each parameter set.  }
\label{tab:directcomparison_scs}
\end{table}

\begin{figure}[t]
\includegraphics[width=1\linewidth]{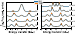}
\caption{ Compton scattering for parameters deviating from the idealized case. All parameters are being fixed except either the inclination angle $\varphi_1$ (left) or the electron kinetic energy $U_e$ (right).}

\label{phase_matching_sweep}
\end{figure}

We now investigate stimulated Compton scattering induced by the interaction of a free-electron wave packet with two inclined laser pulses. Similar to the PINEM effect, when the criterion for the stimulated Compton scattering as specified above is satisfied, a strong modulation of the electron wave packet is observed along the longitudinal direction, leading to the formation of energy sidebands \cite{Tsarev2023}. The entire electron wave packet is deflected as well via the transverse momentum component transfer $\hbar \omega_1\sin(\varphi_1)/c-\hbar\omega_2\sin(\varphi_2)/c$.

Fig.~\ref{fig:time_ev_compton} shows the time evolution of the electron wave packet during interaction with the optical beat-wave. The split-step and the TGWP solution remain in close agreement throughout the interaction process. The TGWP method accurately reproduces the coherent redistribution of the wave function in momentum space. The regularized error remains below one percent during the entire propagation, demonstrating that the phase-space dynamics are captured completely by the mesh-free solution.

The longitudinal energy transfer after the interaction is shown in Fig.~\ref{fig:comparison_compt}. For all investigated parameter sets, the TGWP method reproduces the positions and structure of the momentum sidebands as well as the oscillatory interference structure between the main peaks with high accuracy. 

Compared to the near-field interaction, the stimulated Compton spectra exhibit stronger overlap between neighboring sidebands. As a result, the spectral intensity remains finite over the occupied momentum range and the Gaussian noise in this region does not contribute to the error.

The overall $ \epsilon_{L^2}$ errors are significantly smaller than in the near-field case, ranging from 0.045 to 0.675, increasing linearly with electric field strength. Phase alignment reduces the error considerably with $ \epsilon_{L^2}^\mathrm{abs}$ ranging from 0.018 to 0.129. The TGWP speedup amounts  to $\approx 800$.

To further assess the robustness of the TGWP approxmiation, we investigate parameter regimes that deviate from the ideal phase matching (Eq. \eqref{eq_phase_matching}). Fig.~\ref{phase_matching_sweep} shows the resulting energy transfer spectra when either one of the inclination angles $\varphi_1$ or the electron kinetic energy $U_e$ is varied from its ideal value while all remaining parameters are kept fixed.

As the system is moved away from the optimal phase-matched conditions, the efficiency decreases, resulting in a reduced energy transfer rate and weaker modulation of the wave packet. 
In particular, according to Eq. \eqref{eq_phase_matching}, reducing $\varphi_1$ from $75^{\circ}$ increases the longitudinal momentum component of the optical grating, resulting in a larger separation of momentum sidebands. In contrast, decreasing the electron kinetic energy moves the sidebands closer to the zero-loss peak. Nevertheless, reducing the interaction angle from 75° to 60° does not completely suppress the energy-transfer rate, owing to the finite spatial extent of the optical beams in the interaction region, as in contrast with plane-wave light, which partially preserves the conditions required for stimulated Compton scattering.

Across the varied parameter range, the TGWP solution remains in agreement with the grid-based solution. This demonstrates that the method captures not only the ideal stimulated Compton scattering but also reproduces the physical trends in off-resonant regimes.

\section{\label{sec4} Conclusion}
In summary, we have introduced a mesh-free numerical framework for propagating single-particle electron wave packets through time-dependent electromagnetic potentials by directly solving the Schrödinger equation. The method is based on an overcomplete set of Gaussian wave packets, whose individual evolution in the electromagnetic field can be efficiently tracked and superposed owing to the linearity of the Schrödinger equation. We benchmarked the approach against a previously established grid-based solver \cite{Talebi2020} for two representative cases: (i) a quasistatic plasmonic dipolar near-field and (ii) stimulated Compton scattering arising from two interacting laser pulses in free space. In both scenarios, the mesh-free and grid-based approaches exhibit excellent agreement, while the mesh-free method achieves speedups of up to two orders of magnitude.

Our results establish an efficient and scalable platform for modeling quantum electron-optics experiments beyond classical ray-tracing methods and for simulating free-electron–light interactions beyond the no-recoil approximation.   
TGWP is a promising method for resolving full Maxwell--Schrödinger dynamics, in particular since single thawed Gaussian wave packets have successfully been applied to magnetic (minimal coupling) Hamiltonians \cite{SchBHL25}.
Combined with optimization algorithms and machine-learning techniques, the proposed framework could enable the inverse design of complex electromagnetic environments for the on-demand shaping and coherent control of electron wave packets.\\
\section*{Acknowledgement}

The work of M.~Hochbruck was funded by the Deutsche Forschungsgemeinschaft (DFG, German Research Foundation) -- Project-ID 258734477 –- SFB 1173. 
The work of C.~Lasser was funded by DFG -- Project-ID 470903074 -- TRR 352.
N.~Talebi has received funding from the European Research Council (ERC) under the European Union’s Horizon 2020 research and innovation programme (grant agreements no. 101170341 (Kiel, UltraSpecT), the Deutsche Forschungsgemeinschaft (DFG) under grant numbers 525347396 and 447330010, and the VolkswagenStiftung through a Momentum Grant.

This work was originated at the KIT workshop ``Meshfree Methods for Magnetic Schrödinger-Type Equations'' in October 2025.




\section{Author Contributions}

M.F. and S.M. contributed equally to this work. The \textsc{MATLAB} code was implemented by M.F.and N.T. The \textsc{Julia} code was implemented by S.M. Calculations and postprocessing were done by M.F. and S.M. The manuscript was written by all authors. C.L., M.H., and N.T. provided supervision.

\section{Data Availability}
The data that support the findings of this study are available from the authors upon reasonable request. The TGWP code is openly available at \url{https://gitlab.lrz.de/00000000014AA221/tgwp_meshfree_integrator} .

\bibliographystyle{apsrev4-2}
\bibliography{apssamp}

\end{document}